\documentstyle[aps]{revtex}
\begin{document}
\title{A SPHERICALLY SYMMETRIC CLOSED UNIVERSE
\\ AS AN EXAMPLE OF A 2D DILATONIC MODEL }
\author{L\'{a}szl\'{o} \'{A}. Gergely}
\address{
Laboratoire de Physique Th\'{e}orique, Universit\'{e} Louis Pasteur,
\\ 3-5 rue de l'Universit\'{e} 67084 Strasbourg, France
\\ and
\\ KFKI Research Institute for Particle and Nuclear Physics,
\\ Budapest 114, P.O.Box 49, H-1525 Hungary}
\maketitle
 
\begin{abstract}
We study the two-dimensional (2D) dilatonic model describing a
massless scalar field minimally coupled to the spherically reduced
Einstein-Hilbert gravity. The general solution of this model is given
in the case when a Killing vector is present. When interpreted in
four dimensions, the solution describes either a static or a
homogeneous collision of incoming and outgoing null dust streams
with spherical symmetry. The homogeneous Universe is closed.
\end{abstract}
 
\bigskip
 
\section{Introduction}
 
In a recent paper\cite{GergelyND} we gave the general solution
of the Einstein field equations describing the collision of spherically
symmetric null dust streams, when the resulting space-time is static.
For special parameter values the solution was found independently by
Kramer \cite{KramerND}, who obtained also the static and stationary
solutions describing the collision of  cylindrical null dust beams
\cite{KramerND2}. The spherically symmetric static solution has the
two dimensional interpretation\cite{GergelyND} of a scalar field in
minimal coupling with the dilatonic gravity, which arises from the
dimensional reduction of the Einstein-Hilbert action.
 
In this communication we present the general solution in the case when an
additional Killing vector (not necessarily timelike) is present.
This
corresponds to the $so(3)\oplus R$ algebra in a four dimensional
(4D) picture. The system is characterized by the action
\begin{equation}
S=\int d^{2}x\sqrt{-g}\rho\left[ {\cal R}\left[ g\right] +\frac{1}{2}g^{\alpha
\beta }\nabla _{\alpha }\ln \rho \nabla _{\beta }\ln \rho +\frac{2}{\rho }
\right] -\frac{1}{2}\int d^{2}x\sqrt{-g}g^{\alpha \beta }\nabla _{\alpha
}\varphi \nabla _{\beta }\varphi .  \label{action}
\end{equation}
Here $\rho $ is the dilaton, $g_{\alpha \beta }$ the two-metric, ${\cal R}$
the Ricci scalar of the $g_{\alpha \beta }$-compatible two-connection $
\nabla $ and $\varphi $ is the scalar field. We obtain the dynamical
equations in the following way. First we pass to a conformal metric $\eta
_{\alpha \beta }=h^{-1}g_{\alpha \beta }$ and vary the action (\ref{action})
with respect to $\rho ,\eta ^{\alpha \beta }$ and $\varphi $. (The variation
with respect to $h$ gives nothing new but the trace of the equation obtained
by varying with respect to $\eta ^{\alpha \beta }$.) Then we impose the
flatness of the metric $\eta _{\alpha \beta }$ and choose null coordinates $
x^{\pm }$. In this way the 4D line element takes the form $
ds^{2}=-h(x^{+},x^{-})dx^{+}dx^{-}+\rho (x^{+},x^{-})d\Omega ^{2}$ and the
equations are:
 
\begin{eqnarray}
\delta \varphi &:&\qquad \varphi ,_{+-}=0  \label{dephi} \\
\delta \eta ^{++} &:&\qquad \rho ,_{++}-\rho ,_{+}(\ln \sigma ),_{+}=-\frac{1
}{2}\left( \varphi ,_{+}\right) ^{2}  \label{++} \\
\delta \eta ^{--} &:&\qquad \rho ,_{--}-\rho ,_{-}(\ln \sigma ),_{-}=-\frac{1
}{2}\left( \varphi ,_{-}\right) ^{2}  \label{--} \\
\delta \eta ^{+-} &:&\qquad \rho ,_{+-}+\frac{1}{2}\rho ^{-1/2}\sigma =0
\label{+-} \\
\delta \rho &:&\qquad (\ln \sigma ),_{+-}-\frac{1}{4}\rho ^{-3/2}\sigma =0.
\label{derho}
\end{eqnarray}
Here commas denote derivatives and we have introduced the notation:
$\sigma=h\rho ^{1/2}.$ In deriving Eq. (\ref{derho}) the
trace of the $\delta \eta ^{\alpha \beta }$ equation was employed.
Inserting $\rho ,_{+-}$ from Eq. (\ref{+-}) in the $\partial _{-}$
derivative of Eq. (\ref{++}), we find Eq. (\ref{derho}). This
interdependence of the equations is not surprising, as there are
Bianchi identities to be satisfied. The wave equation (\ref{dephi})
leaves us with the D'Alembert solution $\varphi
=\varphi ^{+}\left( x^{+}\right) +\varphi ^{-}\left( x^{-}\right) $,
a sum of left- and rightmover fields. Before proceeding to solve the
problem in the case when a symmetry is present, we review how the
solution emerges in the cases when no or only one component of the
scalar field is present.
 
\section{Vacuum solution}
 
It is easy to solve the remaining three equations (\ref{++})-(\ref{+-}) in
the vacuum case $\varphi =0$. After dividing by $\rho ,_{+}$ and $\rho ,_{-}$
respectively, the Eqs. (\ref{++}) and (\ref{--}) can immediately be
integrated to obtain
\begin{equation}
\rho _{,\pm }=H^{\mp }\sigma ,  \label{rhopm}
\end{equation}
where $H^{\mp }\left( x^{\mp }\right) $ are arbitrary integration functions
depending only on one coordinate. Next we eliminate $\sigma $ from the above
two equations and (\ref{+-}). The resulting system can be integrated once
more finding
\begin{equation}
H^{\pm }\rho ,_{\pm }=2m^{\pm }-\rho ^{1/2},  \label{rhopm2}
\end{equation}
where $m^{\pm }\left( x^{\pm }\right) $ form a second set of integration
functions. A comparison with (\ref{rhopm}) however leaves us with $m^{\pm
}=M=const$ and the algebraic relation
\begin{equation}
\sigma =\frac{2M-\rho ^{1/2}}{H^{+}H^{-}}.  \label{al1}
\end{equation}
Inserting this expression of $\sigma $ in the Eqs. (\ref{rhopm}) an
integration yields $2M\ln \left| 2M-\rho ^{1/2}\right| +\rho ^{1/2}=K^{\pm
}-F^{\mp }$, where $F^{\mp }\left( x^{\mp }\right) =\int^{x^{\mp }}d\tilde{x}
^{\mp }/2H^{\mp }\left( \tilde{x}^{\mp }\right) $ and $K^{\pm }\left( x^{\pm
}\right) $ are integration functions, which can be again eliminated by
comparing the right hand sides: $K^{\pm }+F^{\pm }=2A=const.$ Thus a second
algebraic relation between $\rho $ and $\sigma $ has emerged:
 
\begin{equation}
\left| 2M-\rho ^{1/2}\right| \exp \left( \frac{\rho ^{1/2}}{2M}\right) =\exp
\left( \frac{2A-F^{+}-F^{-}}{2M}\right) .  \label{al2}
\end{equation}
 
Our choice of the null coordinates $x^{\pm }$ is not unique. They can be
changed by a coordinate transformation belonging to the conformal subgroup
of diffeomorphisms to $\hat{x}^{\pm }=\chi ^{\pm }\left( x^{\pm }\right) $.
Under such a transformation the variable $\rho $ remains unchanged while $
\sigma $ transforms as $\hat{\sigma}=\sigma /\left( \chi ^{+}\right)
,_{+}\left( \chi ^{-}\right) ,_{-}$. By choosing the new null coordinates
defined by the differential equations $4MH^{\pm }\left( \chi ^{\pm }\right)
,_{\pm }+\chi ^{\pm }=0$ and fixing the integration constants in a
convenient way (to annihilate the constant $A$), the Eqs.
(\ref{al1})-(\ref{al2}) become
\begin{eqnarray}
\sigma &=&\frac{16M^{2}\left( 2M-\rho ^{1/2}\right) }{\hat{x}^{+}\hat{x}^{-}}
\label{Schw} \\
\left( 2M-\rho ^{1/2}\right) &\exp &\left( \frac{\rho ^{1/2}}{2M}\right)
=2\left| M\right| \hat{x}^{+}\hat{x}^{-}.  \nonumber
\end{eqnarray}
The remaining freedom in the choice of the null coordinates is to scale one
of them by a constant and the other one by the reciprocal of this constant.
In the second equation of (\ref{Schw}) we have used that the expressions $
\hat{x}^{+}\hat{x}^{-}$ and $2M-\rho ^{1/2}$ have the same sign, as can be
seen from the positiveness of $\sigma $. Inserting $\rho ^{1/2}=R$ and $
\sigma =hR$ we obtain the Schwarzschild solution with mass $M$ (positive or
negative) with the curvature coordinate $R$ and conformal factor $h$ written
in terms of the Kruskal coordinates. A similar derivation was given by Synge
\cite{Synge}.
 
For the value $M=0$ of the first integration constant the derivation has to
be slightly modified. Then in place of Eq. (\ref{al2}) we have $\rho
^{1/2}=2A-F^{+}-F^{-}$ and the new null coordinates $\hat{x}^{\pm }=\pm
2\left( A-F^{\pm }\right) $ are chosen. The solution is the flat space-time
\begin{equation}
h=1,\qquad \hat{x}^{+}-\hat{x}^{-}=2R.  \label{Mink}
\end{equation}
 
\section{Chiral solution}
 
The chiral case, when only one component of the scalar field is present, has
been solved for a quite general class of dilatonic Lagrangians
obtaining generalized
Vaidya solutions \cite{Navarro}. We illustrate how does the solution emerge
for the system (\ref{action}) on the case of the leftmover field $\varphi
=\varphi ^{+}$ (thus $\varphi ^{-}=0$). Then the Eq. (\ref{--}) can be
integrated obtaining Eq. (\ref{rhopm}) containing $\rho_{,-}$. By
inserting $\sigma $ from the Eq. (\ref{+-}) and integrating as in the vacuum
case we find the $(+)$ equation of (\ref{rhopm2}). Inserting this in Eq.
(\ref{++}) and employing again Eq. (\ref{+-}) a first order differential
equation for the function $m^{+}$ emerges, with the solution
\begin{equation}
m^{+}=M-\frac{1}{4}\int dx^{+}H^{+}\left( \varphi ,_{+}\right) ^{2}
\end{equation}
 
We define a convenient null coordinate $V$ by $dV=-dx^{+}/H^{+}.$ (This is
related to the Kruskal coordinate introduced in the vacuum case by $
dV=4Md\ln \hat{x}^{+}.$) Expressed as a function of $V,$ the integration
function $m^{+\text{ }}$becomes
\begin{equation}
m^{+}=M+\frac{1}{4}\int dV\left( \varphi ,_{V}\right) ^{2}.
\label{massVaidya}
\end{equation}
In terms of the curvature coordinate $R$ and conformal factor $h^{\prime
}=-H^{+}h,$ the remaining field equations (the $\rho ,_{-}$ equation from
(\ref{rhopm}) and the $\rho ,_{+}$ equation from (\ref{rhopm2})) give
\begin{equation}
2R,_{-}=-h^{\prime },\qquad 2R,_{V}=1-\frac{2m^{+}(V)}{R} .
\label{RV} \end{equation}
 
Writing the line element (with $h^{\prime }$ and $V$ in place of $h$ and $
x^{+}$) in terms of the coordinates $(R,V)$ we obtain the incoming Vaidya
solution \cite{Vaidya} $ds^{2}=-(1-2m^{+}\left( V\right)
/R)dV^{2}+2dRdV+R^{2}d\Omega ^{2}$, with $m^{+\text{ }}$as the mass
function. Waugh and Lake\cite{WLake} has shown that a closed form of this
solution can be given in double null coordinates only for linear and
exponential mass functions. This implies by Eq. (\ref{massVaidya}) that the
scalar field has to be also a linear or exponential function in order to be
able to integrate the second equation (\ref{RV}).
 
\section{Solution with Symmetry}
 
Until now we have discussed the cases where at least one of the two
components of the scalar field vanishes. If both components are
present,
we can use them in the construction of new null coordinates $\tilde{x}^{\pm
}=\varphi ^{\pm }/\sqrt{2}$. In terms of these null coordinates, dropping
the tildes, we obtain the system:
 
\begin{eqnarray}  \label{gensyst}
\rho ,_{++}-\rho ,_{+}(\ln\sigma ),_{+} &=&-1  \label{pp} \\
\rho ,_{--}-\rho ,_{-}(\ln\sigma ),_{-} &=&-1  \label{nn} \\
2\rho ,_{+-}+\rho ^{-1/2}\sigma &=&0  \label{pn}
\end{eqnarray}
 
The general solution for the system (\ref{pp})-(\ref{pn}) is not yet known
in closed form. (Mikovi\'{c} \cite{Miko} has given the solution in the form
of a perturbative series in powers of the outgoing energy-momentum
component.)
 
Our purpose is to generalize the solution given in \cite{GergelyND} for
any Killing vector tangent to the surface defined by
$\theta=const$ and $\varphi=const.$
From the Killing equations we find that the Killing vector ${\cal
K}
=\left( {\cal K}^{+}\left( x^{+}\right) ,{\cal K}^{-}(x^{-}),0,0\right) $
satisfies ${\cal K}^{+}\rho ,_{+}+{\cal K}^{-}\rho ,_{-}=0$ and $\left(
{\cal K}^{+}h\right) ,_{+}+\left( {\cal K}^{-}h\right) ,_{-}=0.$ The last
relation implies the existence of a potential $N$ defined by $h{\cal K}^{\pm
}=\pm N,_{\mp }$. The remaining Killing equations in terms of the
potential $N$ are:
 
\begin{eqnarray}
&&N,_{-}\rho ,_{+} =N,_{+}\rho ,_{-}
\label{N1}\\
N,_{\pm \pm }&=&N,_{\pm }\left( \frac{\rho ,_{\pm }}{2\rho}
+ \left( \ln \sigma \right),_{\pm } \right) .
\label{N2}
\end{eqnarray}
 
The potential $N$ can be eliminated from the system
(\ref{N1})-(\ref{N2}) in the following way. We express $N_{,+-}$
from the Eq. $\partial_{-}$(\ref{N1}) and we substitute it in the
Eq. $\partial_{+}$(\ref{N1}) in which also $N_{,\pm\pm}$ are
replaced by their expressions (\ref{N2}). By inserting $N_{+}$
from Eq. (\ref{N1}), we obtain a differential equation in $\rho$
and $\sigma$. Eliminating the second derivatives $\rho_{,\pm\pm}$
from Eqs. (\ref{pp}) and (\ref{nn}), we find $\rho ,_{+}=c\rho ,_{-}$
where $c^{2}=1$. This implies that $\rho $
depends on a single variable $x^{+}+cx^{-}$. It is immediate to show that
the same property holds for $\sigma $ and $N$. The Killing vector is ${\cal K
}=\alpha \left( c,-1,0,0\right) $, where $\alpha $ is a constant.
Thus the equations (\ref{pp})-(\ref{pn}) lead to the system of
ordinary differential equations:
\begin{eqnarray}
\frac{d^{2}\rho }{d\left( x^{+}+cx^{-}\right) ^{2}}+\frac{c\sigma }{2\rho
^{1/2}} &=&0  \nonumber \\
\frac{d\rho }{d\left( x^{+}+cx^{-}\right) }\frac{d\ln \sigma }{d\left(
x^{+}+cx^{-}\right) } &=&1-\frac{c\sigma }{2\rho ^{1/2}}.  \label{1var}
\end{eqnarray}
 
We introduce timelike and spacelike coordinates $t$ and $r$ respectively by $
x^{\pm }=t\pm r$ in terms of which the metric is $
ds^{2}=h(-dt^{2}+dr^{2})+\rho d\Omega ^{2}$. There are two cases to discuss:
$\left( a\right) $ when all metric functions depend only on $r$ (static
case) or $\left( b\right) $ when they depend solely on $t$ (homogeneous
case).
 
$\left({\bf a}\right) $ We have discussed this case in detail in
\cite{GergelyND}
. The fact that the dilaton $\rho =R^{2}$ has only radial dependence
suggests to introduce $R$ as a new radial coordinate. The metric becomes $
ds^{2}=-hdt^{2}+f^{-1}dR^{2}+R^{2}d\Omega ^{2}$, where the metric function $
f $ is defined by the differential equation
\begin{equation}
\left( \frac{dR}{dr}\right) ^{2}=fh  \ .\label{fdefstat}
\end{equation}
 
By introducing the notation $\beta =2/h$ we recover from (\ref{1var}) the
equations $(2.7)-(2.8)$ of \cite{GergelyND}.
 
$\left({\bf b}\right) $ Now the dilaton depends only on time. We
introduce $R$ as a new time variable and obtain the metric $
ds^{2}=-f^{-1}dR^{2}+hdr^{2}+R^{2}d\Omega ^{2}$, where the metric
function $ f $ is defined similarly to (\ref{fdefstat}):
\begin{equation}
\left( \frac{dR}{dt}\right) ^{2}=fh  \ .\label{fdefhom}
\end{equation}
From (\ref{1var}) we find equations very much similar to $(2.7)-(2.8)$
of \cite{GergelyND}. Both are written concisely as:
\begin{eqnarray}
f\frac{d\ln f}{d\ln R} &=&-\beta -f-c  \nonumber \\
f\frac{d\ln \beta }{d\ln R} &=&-\beta +f+c\ ,  \label{eqs}
\end{eqnarray}
where $c=-1$ refers to the static case and $c=1$ to the homogeneous case.
 
The solving procedure of this system follows closely \cite {GergelyND},
with the only difference in the definition of
the new variables $P=\pm \left(2f\beta \right)^{1/2}$ and
$L=\pm\left( 1-c\beta \right) /\left( 2f\beta \right) ^{1/2},$
allowed to take either positive or negative values. We find
\begin{eqnarray}
cCR &=&e^{L^{2}}-2L\Phi _{B}(L),\qquad
\Phi_{B}(L)=B+\int^{L}e^{x^{2}}dx  \label{RL} \\
\beta &=&\frac{CR}{e^{L^{2}}},\qquad \qquad f=\frac{2\Phi _{B}^{2}(L)}{
e^{L^{2}}CR}  \label{beL} \\
ds^{2} &=&-\frac{2ce^{L^{2}}R}{C}dL^{2}+\frac{e^{L^{2}}}{cCR}
dr^{2}+R^{2}d\Omega ^{2}  \label{metricrL}
\end{eqnarray}
where $B$ and $C>0$ are constants and Eq. (\ref{RL}) determines
$R$ as function of $L.$
 
\section{Concluding Remarks}
 
The metric (\ref{metricrL}), either static (then $t$ replaces $r$
and $L$ is a radial coordinate) or homogeneous, has a true
singularity at $R=0.$ Although
the Ricci scalar vanishes ${\cal R}=0$, other scalars, like ${\cal R}_{ab}
{\cal R}^{ab}=2C^{2}/e^{2L^{2}}R^{2}$ and the Kretschmann scalar ${\cal R}
_{abcd}{\cal R}^{abcd}$ which has the denominator $\left( CR\right) ^{6}$
are ill-behaved at $R=0.$
 
The 4D interpretation of the source is again twofold. First we have the
general picture of colliding null dust streams, valid even without
the assumption
of a Killing vector. Second we can interpret the source as an anisotropic
fluid with no tangential pressures and both the energy density and radial
pressure equal to $\beta /8\pi R^{2}=C/e^{L^{2}}R$ (these also became
infinite at $R=0$).
 
Note, that in contrast with \cite{GergelyND}, the transcendental
function $\Phi_{B}(L)$ is {\em not} constrained to be positive.
In the static case, because $\Phi_B(L)=-\Phi_{-B}(-L)$, the
negative values of $L$ lead just to an other copy of the solution
written in \cite{GergelyND} for positive $L$, as was explained
in\cite{GergelyNDsal}.
The consequences in the homogeneous case are deeper, as will be seen
in what follows.
A natural requirement the new time variable $L$ should satisfy is
to be a monotonous function of $t$. From (\ref{RL}) the relation
$d(CR)/dL=-2\Phi_B(L)$ emerges. By fixing the sign in the square
root of Eq. (\ref{fdefhom}) appropriately ($-$ for $\Phi >0$ and
$+$ for $\Phi <0$), we get $d(CR)/dt=-2\Phi /R$. In conclusion
\begin{equation}
\frac{dL}{dt}=R>0\ .
\label{ttoL}
\end{equation}
An other remark is that at $d(CR)/dL=0$ the function $f$ vanishes,
thus the metric written in the coordinates $(R,r,\theta,\phi)$ has
a coordinate singularity, which however does not appear in the form
(\ref{metricrL}) of the metric, when the coordinates
$(L,r,\theta,\phi)$ are employed. This feature is closely related
to the fact that both the transformations $t\to R$ and $L\to R$
are ill-behaved at $\Phi_B(L)=0$, whereas the direct
transformation $t\to L$, given by (\ref{ttoL}), is regular.

\begin{figure}[tbh]
\hspace*{.2in}
\special{hscale=30 vscale=30 hoffset=-20.0 voffset=20.0
         angle=-90.0 psfile=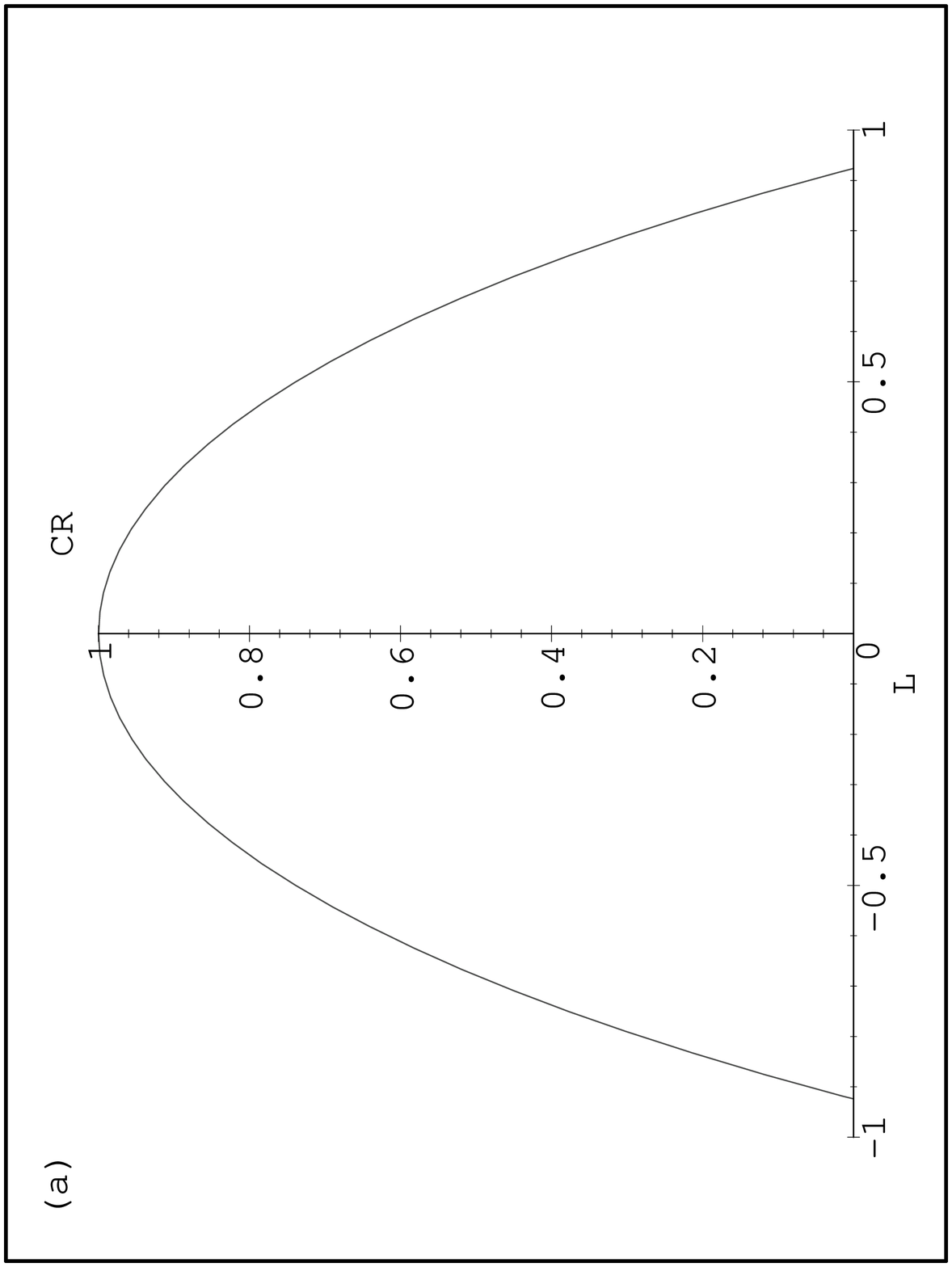}
\hspace*{3.2in}
\special{hscale=30 vscale=30 hoffset=-20.0 voffset=20.0
         angle=-90.0 psfile=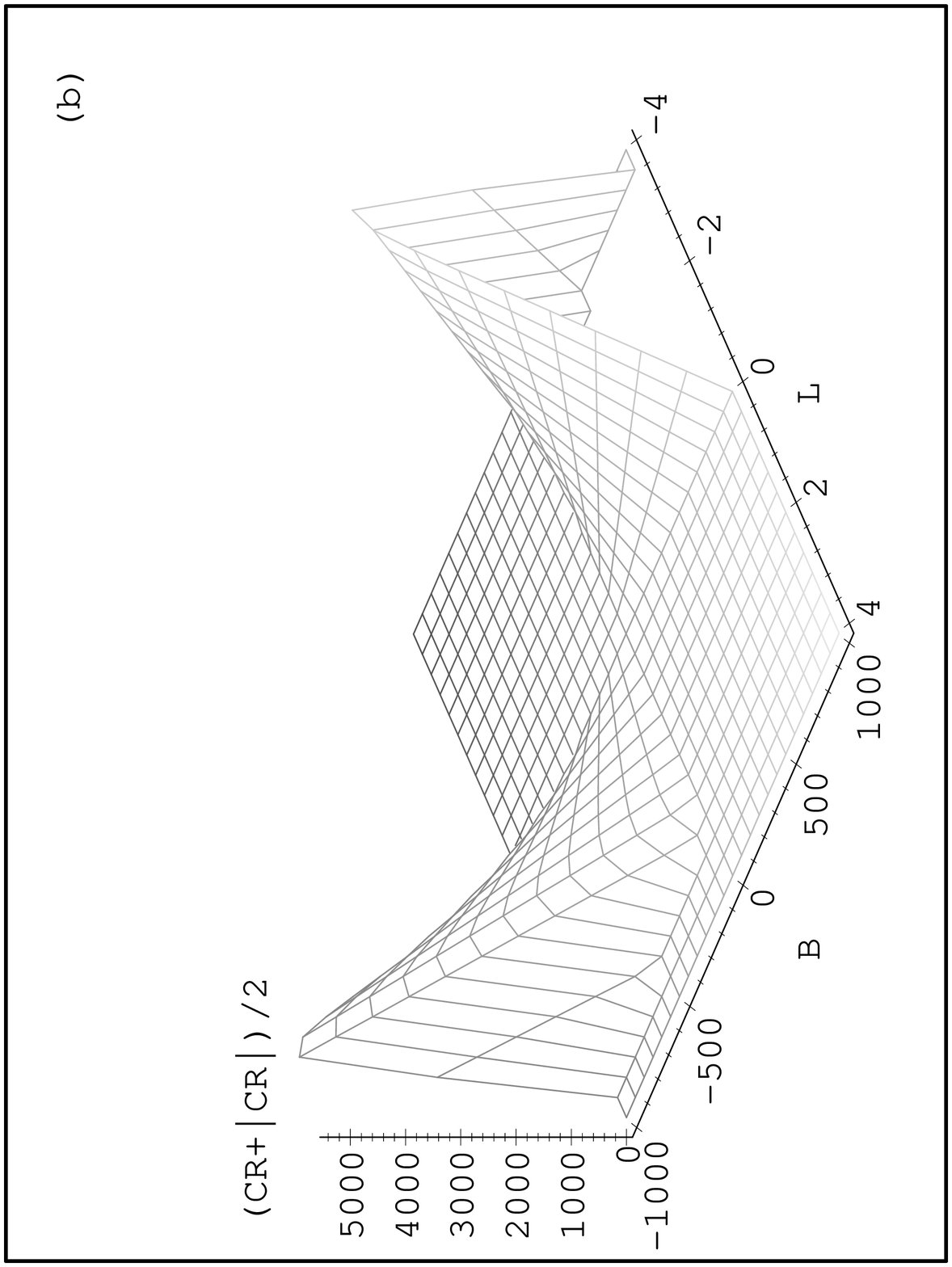}
\vspace*{2.3in}
\caption{(a) The function $CR$ is positive in a domain of $L$
centered on the origin, when the parameter $B=0$ is chosen.
(b) The plot of $(CR+\mid CR\mid )/2$ for a wide range of the
constant $B$ shows a shift of the admissible domain of $L$ in
the positive or negative direction depending on the
sign of $B$ }
\end{figure}
 
A delicate issue is the signature of the metric (\ref{metricrL}).
To have a homogeneous metric $\left( c=1\right) $ with $L$ as time
and $r$ as radial coordinate, as claimed, the condition $R>0$ should
be fulfilled. This translates to have $L$ confined to a finite range
$L\in \left(L_{0-},L_{0+}\right) ,$ as can be seen on Fig.1. from
the numerical plot of (\ref{RL}). The constant $C$ provides
a scale as in the static case. The other constant
$B$ shifts the admissible domain of $L$ to negative values when $B>0$
and conversely. The singularity is on the boundaries
$L_{0\pm}$ of the admissible range of $L.$
 
The function $R$ is the time-dependent radius of the Kantowski-Sachs
type homogeneous Universe described by (\ref{metricrL}). It is first
increasing to a maximum value after which it decreases to zero. This
Universe filled by a two-component radiation is born from a singularity
and collapses into an other singularity.
 
\section*{Acknowledgments}
 
The author wishes to thank Z. Perj\'es for comments on the
manuscript, M. Bradley for a remark and for the warm hospitality
encountered in the relativity group at the University of
Utah, where the first part of this research was completed.
This work has been supported by the Hungarian State E\"{o}tv\"{o}s
Fellowship and OTKA grant no. D23744. The algebraic packages
REDUCE and MapleV were used for checking computations and
numerical plots.

\end{document}